\documentclass[12pt,a4paper]{article}
 
\usepackage{amsmath, amssymb, mathrsfs, graphicx,subcaption, dsfont, enumerate, epstopdf, xcolor, slashed}
\usepackage{cite} 
\usepackage[colorlinks, allcolors=blue!70!black, linktocpage]{hyperref}

\numberwithin{equation}{section}

\def\half{\frac{1}{2}}
\def\eq#1 { \begin{equation} #1 \end{equation} }
\def\eqn#1{ \begin{align} #1 \end{align} }

\def\a{\alpha}

\def\g{\gamma}
\def\e{\epsilon}

\def\w{\omega}

\def\s{\sigma}

\def\d{\partial}
\def\cL{\mathcal{L}}

\def\cA{\mathcal{A}}
\def\cF{\mathcal{F}}
\def\cU{\mathcal{U}}

\def\cO{\mathcal{O}}

\def\cB{\mathcal{B}}

\def\sl2r{SL(2,\mathbb{R})}

\def\nablat{{\tilde \nabla}}

\def\wt{{\tilde \omega}}

\def\ce{\varepsilon}

\newcommand{\df}[1]{\boldsymbol{#1}}

\newcommand{\lsim}{\mathrel{\hbox{\rlap{\lower.55ex \hbox{$\sim$}} \kern-.3em \raise.4ex \hbox{$<$}}}}
\newcommand{\gsim}{\mathrel{\hbox{\rlap{\lower.55ex \hbox{$\sim$}} \kern-.3em \raise.4ex \hbox{$>$}}}}

 \newcommand{\be}{\begin{equation}}
\newcommand{\ee}{\end{equation}}

\usepackage{bm}

\newcommand{\lrvec}[1]{\stackrel{\leftrightarrow}#1}

\newcommand{\vf}{v_{\scriptscriptstyle{\rm F}}}

\begin{document}

\title{\begin{flushright}\vspace{-1in}
       \mbox{\normalsize  EFI-17-20}
       \end{flushright}
       \vskip 20pt
Electrons and  composite Dirac fermions \\in the lowest Landau level
}

\date{\today}

\author{
	Kartik Prabhu\thanks{\href{mailto:kartikprabhu@cornell.edu}         {kartikprabhu@cornell.edu}} \\               {\em \it Cornell Laboratory for Accelerator-based Sciences and Education (CLASSE)}\\
   {\it   Cornell University, Ithaca, NY 14853, USA}
    \and 
	Matthew M. Roberts\thanks{\href{mailto:matthewroberts@uchicago.edu}
     {matthewroberts@uchicago.edu}} \\
 {\em  \it  Kadanoff Center for Theoretical Physics}\\
   {\it   University of Chicago, Chicago, IL 60637, USA}
} 

\maketitle

\begin{abstract}

We construct an action for the composite Dirac fermion consistent with symmetries of electrons projected to the lowest Landau level. First we construct a generalization of the $g=2$ electron that gives a smooth massless limit on any curved background. Using the symmetries of the microscopic electron theory in this massless limit we find a number of constraints on any low-energy effective theory. We find that any low-energy description must couple to a geometry which exhibits nontrivial curvature even on flat space-times. Any composite fermion must have an electric dipole moment proportional and orthogonal to the composite fermion's wavevector. We construct the effective action for the composite Dirac fermion and calculate the physical stress tensor and current operators for this theory.

\end{abstract}

\newpage

\section{Introduction}\label{sec:intro}

Interacting electrons in a magnetic field can form strongly entangled incompressible fluids: the fractional quantum Hall states  \cite{Tsui:1982yy,Laughlin:1983fy}. One of many interesting families of fractional quantum Hall states are the Jain states, with filling fractions $\nu = \frac{N}{2N+1},~\frac{N+1}{2N+1}$ respectively, which have been observed from $N=1$ all the way to $N=10$ \cite{PhysRevB.77.075307}. These states are crucially related to the theory of the half-filled Landau level, a gapless state conjectured to be described by a neutral Fermi liquid coupled to a dynamical gauge field. These fermions have been observed experimentally \cite{Willett:1997cr,PhysRevLett.71.3850,PhysRevLett.72.2065}. An early conjectured theory of the half-filled state is the Halperin-Lee-Read theory \cite{PhysRevB.47.7312}. This description, however, fails to capture all of the features of electrons projected to the lowest Landau level and the particle-hole symmetry of electron-electron interactions in this limit \cite{PhysRevX.7.031029,GNRS_to_appear}. A new conjectured model for electrons near half filling is Son's composite Dirac fermion model \cite{Son:2015xqa,Son:2016zpg}. Its key ingredients are a neutral Fermi surface\footnote{The original HLR theory also has composite fermions which are electrically neutral, as emphasized by Read \cite{Read:1994yg,READ19967}.} with Berry phase $\pi$ coupled to a dynamical gauge field. This theory is, on flat space and to lowest order in derivatives,\footnote{While the Chern-Simons terms appear to be incorrectly quantized, this can be fixed by a field redefinition $a \rightarrow -2a$ and adding an additional gauge field as Lagrange multiplier as discussed in \cite{Seiberg:2016gmd}.}
\eq{\label{eq:trivial_flat}
i \overline \Psi \gamma^\mu\left( \d_\mu - i  a_\mu \right) \Psi - \frac{1}{4\pi} Ada+\frac{1}{8\pi} AdA .
}
Here  $\Psi$ is the composite Dirac fermion, as opposed to the original electrons which we will designate $\Psi_e$, and our  conventions for Chern-Simons terms are that $Ada = \ce^{\mu\nu\rho} A_\mu \d_\nu a_\rho$. In this paper we will use the representation
\eqn{\label{eq:gamma_rep_pauli}
\gamma^0 = \s^3,~\g^1 = i \s^2,~ \g^2 = - i \s^1,\nonumber \\
i \overline\Psi \g^\mu D_\mu \Psi = i \Psi^\dagger D_0 \Psi + i \Psi^\dagger \s^i D_i \Psi.
}
Note that as the gauge field $a$ is linear in the action it acts as a Lagrange multiplier enforcing the constraint that
\eq{
\Psi^\dagger \Psi = j_\mathrm{CF}^0 = \frac{B}{4\pi},
}
so the  magnetic field acts as a chemical potential, giving us a Fermi surface. geometric fluctuations of this Fermi surface can be identified with the magneto-roton density wave excitations of some fractional quantum Hall states \cite{PhysRevLett.117.216403}. We can calculate the total electric charge as well by varying $A_0$, giving
\eq{
j^0 =  \frac{B-b}{4\pi}
}
where $b$ is the magnetic field of $a$. To move away from half-filling we must turn on a magnetic field in $a$. We use a regularization convention for the Dirac field that is particle-hole symmetric, so that $N_{CF}=0$ when we fill all hole levels and half of the zeroth level. The degeneracy of each Dirac Landau level on the plane is $|n_\phi|$, where $n_\phi = \int \frac{b}{2\pi}$. Therefore if we tune $n_\phi = \pm \frac{N_\phi}{2N+1}$ we have precisely enough composite fermions to fill $N$ positive Landau levels. We can easily calculate the filling fraction to be $\nu = \frac 12 \mp \frac{1}{4N+2}$, the Jain series.

In such a state, integrating out the composite fermions generates the Chern-Simons term \eq{
\pm\frac{N+\half}{4\pi} ada}
where the $\pm$ is given by the sign of $b$. Ignoring higher derivative terms and integrating out $a$ gives the Chern-Simons term \eq{
\left(\frac 12 \mp \frac{1}{4N+2} \right)
\frac{1}{4\pi}AdA
=
\begin{cases}
	\frac{N}{2N+1}\frac{1}{4\pi}AdA &\text{for~} b>0 \\[1.5ex]
	\frac{N+1}{2N+1}\frac{1}{4\pi}AdA &\text{for~} b<0
\end{cases}
}
an effective action which of course gives the correct Hall conductance and filling fraction for the Jain sequences. 

Consider now the composite fermion on a sphere with flux $n_\phi$ so that we can calculate the topological shift \cite{Wen:1992ej}. We need to couple the Dirac field to the spin connection as well as add additional Chern-Simons terms,
\eq{\label{eq:son_curved}
i\overline\Psi \g^\mu (\d_\mu - i a_\mu + \frac{i}{2}\sigma^3 \w_\mu)\Psi - \frac{1}{4\pi} (A+\half \w)da 
 + \frac{1}{8\pi} (A+\half \w)d(A+\half\w).\footnote{One can motivate the combination $A+\half\w$ that we couple to by starting with the relativistic particle-vortex duality \cite{Seiberg:2016gmd} between two relativistic spin $1/2$ theories and then the transformation property that allows us to relate spinful and spinless electron theories \eqref{eq:spin_shift}, though it is equally justified by reproducing the correct topological shift for spinless electrons in the lowest Landau level.}\footnote{There are also additional gravitational Chern-Simons terms $\Gamma d\Gamma + \frac 23 \Gamma^3$ and $\w d\w$ that we will not keep track of.}} The number of fermions it takes to fill up to the $N$\textsuperscript{th} positive Dirac Landau level on a sphere with flux $n_\phi$ is 
  \eq{
 |n_\phi| (N+1/2) + N(N+1)
 }
so upon integrating out the composite fermion we must generate the following Chern-Simons terms
\eq{
\pm \frac{N+\half}{4\pi}ada + \frac{N(N+1)}{4\pi}ad\w.
}
Integrating out $ a$ now gives
\eqn{
\frac{N}{2N+1}\frac{1}{4\pi} Ad(A + (N+2) \w) &\quad\text{~for~} n_\phi>0, \\
\frac{N+1}{2N+1}\frac{1}{4\pi} Ad(A + (1-N) \w) &\quad\text{~for~} n_\phi<0,
}
confirming that the shift is $N+2$ or $1-N$, in agreement with particle-hole symmetry \cite{PhysRevB.90.115139,PhysRevLett.114.016805}. The main goal of this paper is to generalize \eqref{eq:son_curved} to general curved non-relativistic space-times while enforcing symmetries of  lowest Landau level electron theories.\\

The rest of this paper is structured as follows. In section \ref{sec:LLLelectron} we define a generalization of the $g=2$ electron such that it has a smooth massless limit on arbitrary curved backgrounds. We find that this $g=2,~m=0$ theory has symmetries in addition to Galilean invariance, and discuss how these symmetries constrain any low-energy effective theory. These constraints forbid a neutral particle from coupling directly to the original spin connection, and we must therefore define a new connection to which any low-energy description must couple. In section \ref{sec:composite-fermion} we construct an action for the Dirac composite fermion which at lowest order in derivatives gives \eqref{eq:son_curved}, finding that including the time derivative necessitates also including an electric dipole moment of the form $\vec{d} = \ell_B^2 \hat z \times \vec k$ where $\ell_B = 1/\sqrt{B}$ is the magnetic length..

A note on conventions: as we wish to couple our theories to general non-relativistic curved backgrounds and will use the Newton-Cartan conventions outlined in \cite{Geracie:2015dea,Geracie:2015xfa}, which also contain extensive references to the vast literature on Newton-Cartan geometry. For a quicker review on this formalism see the review in \cite{Geracie:2016dpu}.  However we stress a notational difference, in this paper the Newtonian gravitational potential form is denoted $\df \a$, not to be confused with the statistical gauge field of the composite fermion theory $\df a$. 

\section{Lowest Landau level electrons in \(2+1\) dimensions}\label{sec:LLLelectron}

Projection to the lowest Landau level is only well-defined for general electromagnetic field backgrounds when the electron has a magnetic dipole moment with $g=2$. It is useful to recall that at $g=2$ we can rewrite the (flat space) Schr\"odinger action for an electron $\psi_e$ via Hubbard-Stratonovich transformation \cite{Geracie:2014nka} as
\eq{
\cL = \frac i2 \psi^\dagger_e\!\lrvec{D}_t\!\psi_e - \chi_e(D \psi_e^\dagger ) - \chi_e^\dagger \overline D \psi_e + m \chi_e^\dagger \chi_e + \cL_{int}(\psi_e,\psi_e^\dagger).	\label{eq:g2_flat_action}
}
where $D_\mu = \d_\mu -i A_\mu,~D = D_1 - i D_2,~\overline D = D_1 + i D_2$. It is clear from this construction that as $m \rightarrow 0$ the $\chi_e$ equation of motion is the holomorphic constraint $\overline D \psi_e=0.$ This is simply the second quantized version of the holomorphic constraint on the many-body wavefunction. Holomorphic trial wavefunctions have been used to calculate effective actions \cite{PhysRevB.91.165306,Can:2014ota,PhysRevLett.118.206602}. It has also explicitly been shown that the integer quantum Hall effective action is regular in this limit \cite{Abanov:2014ula,Nguyen:2016itg}. 

Recalling that the nonrelativistic limit of a Dirac electron gives a nonrelativistic theory with $g=2$, we will consider Dirac-like representations similar to the 3+1D representations discussed in \cite{LevyLeblond:1967zz,Montigny}. Consider the following representation of $Gal(2,1)$,
\eq{\label{eq:g_2_gal}
K_e^1 = \begin{pmatrix} 0 & 0 \\ i/2 & 0\end{pmatrix},~
K_e^2 = \begin{pmatrix} 0 & 0 \\ -1/2 & 0\end{pmatrix},~
J_e = \begin{pmatrix} s & 0 \\ 0 & s-1 \end{pmatrix},}
\eq{
\Lambda_e = \begin{pmatrix} e^{is\theta} & 0 \\
-\half e^{is\theta}  (k_1 + i k_2)& e^{i(s-1)\theta} \end{pmatrix},~\Psi_e \rightarrow \Lambda_e\cdot \Psi_e,~ \Psi_e = \begin{pmatrix} \psi_e \\ \chi_e \end{pmatrix}.
}
As we will demonstrate, this describes an electron of spin $s$ and a $g$-factor of 2.

The connection in this basis is
\eq{
\df \w_e =i J_e \df \w - i K_e^a \df \varpi^a =\begin{pmatrix}
i s \df \w & 0 \\
\half \left(\df \varpi^1+ i \df \varpi^2 \right) & i (s-1) \df \w
\end{pmatrix},
}
such that the covariant derivative
\eq{
\nabla \Psi_e = \d  \Psi + \df \w_e \cdot \Psi_e
}
transforms covariantly, $\nabla \Psi_e \rightarrow \Lambda_e \cdot \nabla\Psi_e$. We can similarly construct the transformation properties for the conjugate field $\Psi_e^\dagger$,
\eq{
\Lambda_{e^\dagger} = (\Lambda_e)^\dagger,
\df \w_{e^\dagger} = (\df \w_e)^\dagger
}
\eq{
\Psi_e^\dagger \rightarrow \Psi_e^\dagger (\Lambda_e)^\dagger,~
\nabla \Psi_e^\dagger = \d  \Psi_e^\dagger +  \Psi_e\cdot (\df \w_e)^\dagger.
}
As with  single component Schr\"odinger representations, if this field is massive then due to the transformation properties of $\df \a$ we must construct a covariant extended derivative,
\eq{
D_\mu \Psi_e = \nabla_\mu \Psi_e -i A_\mu \Psi  -i m \a_\mu \Psi_e , D_A \Psi_e = e^\mu{}_A D_\mu \Psi_e,~ D_I\Psi_e = (D_0\Psi_e, D_a\Psi_e, i m \Psi_e),
}
which transforms as $D_I \Psi_e \rightarrow (\Lambda^{-1})^J{}_I \Lambda_e D_J \Psi_e$.
Following the intuition of the one-derivative theories in 3+1D described in \cite{LevyLeblond:1967zz,Montigny} we wish to write a Dirac-like action. There exists a set of invariant matrices,
\eq{
\g^I = 
\begin{pmatrix}
 \begin{pmatrix}1 & 0 \\ 0 & 0 \end{pmatrix} \\\s^1 \\ \s^2 \\\begin{pmatrix}0 & 0 \\ 0 & -2 \end{pmatrix} 
 \end{pmatrix}
 ,~\Lambda_e^\dagger \g^I \Lambda_e = \Lambda^I{}_J \g^J,
}

such that the kinetic term
\eq{
\frac i2 \left( \Psi_e^\dagger \g^I D_I \Psi_e - D_I \Psi_e^\dagger \g^I \Psi_e \right)
}
is invariant under local Galilean transformations and reduces to \eqref{eq:g2_flat_action} in flat space. Note that $\chi_e$ has no time derivatives\footnote{This is only true on causal backgrounds: when $\df n \wedge d \df n=0$  we can  define a global time coordinate $t$ such that $n = f(t,x,y) dt$ and the action has no $t$ derivatives of $\chi_e$.} and is simply a Lagrange multiplier. We can integrate it out and find a Schr\"odinger action with a $g$-factor with $g=2$, 
which gives the equation of motion for $\psi_e$
\eq{\label{eq:el_g_eom}
\frac{1}{2m}g^{IJ} (D_I -\half T_I) (D_J - \half T_J) \psi_e +\frac{i}{4m} n_A \ce^{ABC}\hat{R}_{AB} \psi_e=0
}
where $T_I = \Pi^A{}_I e^\mu{}_A T^\lambda{}_{\lambda \mu}$ and  $\hat{R}_{AB}$ is a generalized curvature tensor (as we will see this is really only purely a curvature tensor on torsionless backgrounds) defined as follows: Recall that for a derivative operator we define the torsion and curvature as
\eq{
(D_I D_J - D_J D_I)\psi_e = R_{IJ}\psi_e - T^K{}_{IJ} D_K \psi_e.
}
This curvature term has suppressed spinor indices and is not simply an index contraction of the Riemann tensor. The still messier object $\hat{R}_{IJ}\psi_e$ defined as
\eqn{
\hat{R}_{IJ} \psi_e &= \left[ (D_I - \half T_I) (D_J - \half T_J)-(D_J - \half T_J) (D_I - \half T_I )\right]\psi_e \\
&=R_{IJ}\psi_e - T^K{}_{IJ}D_K \psi_e - \half \left(D_I T_J - D_J T_I \right) \psi_e
}
is $n$-orthogonal in both indices, and therefore can be written as a pushforward
\eq{
\hat{R}_{IJ} = \Pi^A{}_I \Pi^B{}_J\hat{R}_{AB}.
}
The second term in \eqref{eq:el_g_eom} is a generalized $g=2$ magnetic moment term containing $\frac{B}{2m}\psi_e - \frac{s R}{2m}\psi_e$ among others. It is clear from this construction that integrating out $\chi_e$ to find a standard Schr\"odinger action does not work at $m=0$. The $\chi_e$ equation of motion is
\eq{
2 m i \chi = \left(D_1 + i D_2\right)\psi_e - \half \left(T_1 + i T_2 \right)\psi_e.
}
At $m=0$ (on torsionless backgrounds) the $\chi$ equation of motion is the curved space generalization of lowest Landau level constraint that the wave-function is holomorphic, $\bar D \psi_e =0$. We therefore identify our two-component theory at $m=0$ as the covariant generalization of the $g=2,~m=0$ limit of electrons in a magnetic field\footnote{Earlier attempts at constructing effective actions in the $g=2$ massless limit can be found in \cite{Son:2013,PhysRevLett.113.266802}}. 

Note that in the massless limit we do not need to work with the extended derivative $D_I \Psi_e$ and can write the action simply as
\eq{
\frac i2 \left( \Psi_e^\dagger \g^A D_A \Psi_e - D_A \Psi_e^\dagger \g^A \Psi_e\right),
}
where $D_A \Psi_e= e_A^\mu \left(\d_\mu + (\w_L)_\mu -i A_\mu  \right)\Psi_e$ and $\g^A = \Pi^A{}_I \g^I = \left(\begin{pmatrix}1 & 0 \\ 0 & 0\end{pmatrix}, \s^1,\s^2 \right)$.

We can also include Coulomb interactions as the electron density $\Psi_e^\dagger n_A \sigma^A \Psi_e = \psi_e^\dagger \psi_e$ is Galilean invariant, and so we can include for instance a local $|\psi_e|^4$  interaction.\footnote{It is possible to include other long-range interactions though they can be more difficult, as our formulation is intrinsically three dimensional. For instance to include a $1/r$ Coulomb interaction this is most easily done by confining the electrons to a 2+1D hypersurface and allowing the Coulomb potential to exist in 3+1D, such that taking $\rho \phi + \phi \Box_3 \phi$ leads to a $1/r$ potential for density-density interactions.}

\subsection{Symmetries of the microscopic theory}\label{sec:symm}

The $g=2$ electron has more symmetries than just local Galilean invariance. Unlike the case of a single component nonrelativistic theory, where one can have a symmetry exchanging $A,~\w,~\a$, we have a similar symmetry mixing their temporal components. In particular, the action is invariant under
\eqn{
\df A & \rightarrow \df A + (s \xi + \half \e^{ab} \zeta_{ab}- m \theta) \df n, \nonumber \\
\df \w & \rightarrow \df \w + \xi \df n, \nonumber \\
\df \a & \rightarrow \df \a + \theta \df n, \nonumber \\
\df \varpi^a & \rightarrow \df \varpi^a + \zeta^a{}_b \df e^b.
}
Assuming these shift symmetries are not anomalous, general coordinate covariance requires that the generating functional must be of the form
\eq{\label{eq:gen_sym}
Z\left[\df e^I, \df \omega^A{}_B, \df A\right] = Z\left[\df e^A, \df A +m \df \a - s \df \w - \tfrac{1}{2} \Omega \df n\right]
}
where
\eq{
\Omega = u^\mu \ce_{\mu\nu\rho} h^{\nu\lambda} \nabla_\lambda u^\rho = \d_1 u^2 - \d_2 u^1 + \varpi^2{}_1 - \varpi^1{}_2 - \w_a u^a
} is the vorticity of an as yet unspecified unit normal vector (that is, $n_\mu u^\mu = +1$).

We can relate two different theories with differing electron spin by a field redefinition, in particular
\eq{
Z[s', \df A, \df \w] = Z[s, \df A + (s-s') \df \w, \df \w].\label{eq:spin_shift}
}
Because of this for the rest of this work we will assume the electron to have spin zero with respect to $SO(2)$, knowing that we can easily shift to arbitrary values.

Taking the LLL limit at $s=0$ yields an even more constrained theory, where the partition function must be of the form
\eq{\label{eq:LLL_sym}
Z[\df e^A, \df A - \tfrac{1}{2} \Omega \df n]
}
which of course implies
\eq{
\frac{\delta Z}{\delta \a_\mu}|_{\delta \df \w^A{}_B =0} = 0,\label{eq:no_mass}
}
implying the (unimproved) mass current vanishes identically. 

There are a few additional constraints: If we treat $ \omega_C{}^A{}_B = e_C^\mu\w_\mu{}^A{}_B$ as the variable independant of the tetrad, then\footnote{If we chose $s=\half$ then this changes to $\frac{\delta Z}{\delta \w_1} = 
\frac{\delta Z}{\delta \w_2} = 
\frac{\delta Z}{\delta \varpi^1{}_0} = 
\frac{\delta Z}{\delta  \varpi^1{}_1} = 
\frac{\delta Z}{\delta  \varpi^2{}_0} = 
\frac{\delta Z}{\delta  \varpi^2{}_2} = 0$.}
\eq{\label{eq:connection_sym}
\frac{\delta Z}{\delta \w_0} = 
\frac{\delta Z}{\delta \varpi^1{}_0} = 
\frac{\delta Z}{\delta  \varpi^1{}_1} = 
\frac{\delta Z}{\delta  \varpi^2{}_0} = 
\frac{\delta Z}{\delta  \varpi^2{}_2} = 0.}

In what follows we will focus on the spinless LLL electron theory and thus define \eq{
\df \cA = \df A - \half \Omega \df n,
}
where the choice of unit timelike vector will be described in the next section.

\subsection{The effective drift velocity }\label{subsec:drift}

As stated above we need a timelike vector whose vorticity we use to construct $\df \cA$. 
Following the philosophy of \cite{Geracie:2016inm}, we will construct an effective drift velocity from $\df \cA$ and equate that with the $u^A$ used in defining the vorticity $\Omega$,
\eqn{\label{eq:_self_cons}
\df \cF = d\df \cA,~ \cB = n_\mu \ce^{\mu\nu\rho} \cF_{\nu\rho}/2, \nonumber \\
\cU^\mu = \ce^{\mu\nu\rho} \cF_{\nu\rho}/2 \cB,~ u^\mu = \cU^\mu.
}
We can formally solve this in a derivative expansion, as $u$ is two derivatives higher than $\df A$. The leading term is simply the drift velocity of the electromagnetic field,
\eq{
u^\mu = \ce^{\mu\nu\rho}F_{\nu\rho}/2B + \ldots
}
If we consider small perturbations about flat space with a constant magnetic field $B_0$, we find
\eq{
\delta u^a(k) = \frac{
2 \delta e^a{}_t + \ell_B^2 k^a k^b\delta e^b{}_t + 2 \ell_B^2 \ce^{ab} \delta E^b + \ell_B^4 k^a \w \delta B + i \ell_B^2 \ce^{ab}k^b (\delta \varpi^1{}_y- \delta \varpi^2{}_x)
}{
2 + \ell_B^2\vec{k}^2
},
}
where $\ell_B = 1/\sqrt{B_0}$ is the magnetic length. which at leading order in derivatives is
\eq{
\delta u^a = \delta e^a{}_t +\ell_B^2 \ce^{ab}\delta E^b+ \cO(k)
} 
as claimed above. Note that if we restrict to torsion-free backgrounds, then we are guaranteed that $\df \omega^A{}_B \sim \d \df e^I$, and therefore
\eq{\label{eq:drift_lowest}
u^\mu = \ce^{\mu\nu\rho} F_{\nu\rho}/2B + \cO(\d^2).
}
Note that while we could have added additional higher derivative terms to the definition of $\df \cA$ or $\cU$, at lowest order any construction of a preferred frame consistent with the symmetries described will reduce to \eqref{eq:drift_lowest} at lowest order.

\subsection{Effective connection}\label{sec:eff_conn}
Our conjectured composite Dirac fermion theory \eqref{eq:son_curved} couples to a spin connection in order to reproduce the topological shift. Due to \eqref{eq:connection_sym} we cannot couple any low-energy theory directly to $\df \w$.\footnote{Even if we consider the case of spinful electrons, we find that the spin connection must appear in combination with the electromagnetic field as $\df A - s \df \w$,  and therefore neutral particles cannot minimally couple to $\df\w$.} Unlike in Lorentzian systems, there is no preferred torsionless backgrounds except in the special case where $d\df n=0$. However, even in this case, the structure equation defining the Newtonian connection,
\eq{
\df T^I = d\df e^I + \df \w^I{}_J \wedge \df e^J = 0
}
yields a connection dependent on $\df \a$, violating \eqref{eq:no_mass}. We instead need to choose a structure equation defining a new connection $\df \wt$ satisfying our constraints on the LLL electron theory. Consider the connection defined by the following structure equation\footnote{For alternative methods of treating torsion in transport of topological phases see for example \cite{Bradlyn:2014wla,Gromov:2014vla}},
\eq{\label{eq:comp_structure}
\df T^I(\df e^I, \df \wt^A{}_B) = u^I d\df n + n^I d\overset u {\df \a},
}
where $\overset u {\df \a} = u_I \df e^I = \df \a + \df e^a u^a - \frac 12 u_a u^a \df n$ is simply $\df \a$ measured in the frame $e^\mu{}_0 = u^\mu$ and $u^I = (1,u^a, - \frac 12 u_au^a)^T$ is the null extension of $u^A$. Direct inspection of components confirms that $\df \a$ cancels from both sides of the lowest component equation,
\eq{
T^4 = d \df \a - \df \varpi^a \wedge \df e^a = d(\df \a + \df e^a u^a) - u_a u^a d\df n - \frac 12 d(u_a u^a)\wedge \df n
}
 and therefore this connection is independent of $\df \a$. Even on backgrounds where $d \df n=0$ this connection is explicitly different than the Newtonian connection, as it  does not carry information about the gravitational potential $\df \a$.

In flat space, the connection is
\eq{
\wt_t = \tilde\varpi^1{}_y  = - \tilde \varpi^2{}_x =\half\nabla\times u,~\tilde\varpi^a{}_t = -\frac 12 \d_a (\vec u^2)-\d_t u^a,
}
with all other components zero. The most important result from this is that for small electromagnetic perturbations about a constant magnetic field we have
\eq{
\df \wt = -\frac{ \nabla \cdot E}{2B} dt+\cO(\delta^2),~\tilde{\df\varpi}^i = - \frac{\nabla \cdot E}{2B}\ce^{ij}dx^i - \frac{\ce^{ij} \dot{E}^j}{B} dt+\cO(\delta^2).
}
This strange looking connection has a reasonably simple geometric interpretation, which we can study by looking at the geodesic equation. In flat space,
\eq{
v^\nu \tilde\nabla_\nu v^\mu =0
}
where we emphasize $v^\mu$ is a velocity field for our gedesics and \emph{not} a Newton-Cartan field, 
can be rewritten as a geodesic coupled to external effective electromagnetic fields
\eq{
\dot{v}^i + v^j\d_j v^i = F_\mathrm{(eff)}{}^i{}_j v^j ,~E_\mathrm{(eff)}^i = - \dot{u}^i  - \half \d_i(\vec{u}^2),~ B_\mathrm{(eff)} = \nabla\times u.
}
In the limit of small perturbations about a constant magnetic field, this is simply
\eq{
E^i_\mathrm{(eff)} = - \frac{\ce^{ij} \dot{E}^j}{B},~ B_\mathrm{(eff)} = - \frac{\nabla\cdot E}{B}.
}
Note that, from our construction, \emph{any} effective theory describing electrons in the lowest Landau level must couple to this connection. An interesting question we do not investigate further here is whether there is a deeper physical interpretation of this geometry.

\section{Dirac composite fermion}\label{sec:composite-fermion}

We wish to write a composite Dirac fermion action in a way respecting the symmetries of our $g=2,~m=0$ electron. Recall that in flat space 2+1D Dirac kinetic term is
\eq{
i\overline{\Psi} \gamma^\mu \d_\mu \Psi  = i \Psi^\dagger \left[ (\gamma^0)^2 \d_0 + \g^0 \g^i \d_i \right] \Psi = i \Psi^\dagger \left[ \d_0 + \sigma^i \d_i \right] \Psi 
}
in the representation $\gamma^0 = \s^3, ~ \g^1 = i \s^2, \g^2 = - i \s^1.$

Up to field redefinitions there are only two different two-dimensional representations for fixed spin (that allow for gamma matrix kinetic terms) of $Gal(2,1)$, namely \eqref{eq:g_2_gal} and
\eq{
J = \begin{pmatrix} s & 0 \\ 0 & s-1 \end{pmatrix},~K^a = \begin{pmatrix} 0 & 0 \\ 0 & 0 \end{pmatrix}.
}
The composite Dirac fermion has $s=\half$.\footnote{Unlike in Lorentzian systems we are not required to take $s=1/2$, however this choice is necessary to reproduce the correct shift of Jain states.} In this representation
\eq{\label{eq:dirac_trans}
\Lambda = \begin{pmatrix}e^{i\theta/2} & 0 \\ 0 & e^{-i\theta/2} \end{pmatrix},~\Psi \rightarrow \Lambda \cdot \Psi, ~ \Psi = \begin{pmatrix} \psi \\ \chi \end{pmatrix}.
}
Enforcing \eqref{eq:no_mass} means that the composite fermion is massless, $M\Psi=0.$ The covariant derivative of $\Psi$ is
\eq{
\nablat_\mu \Psi = \d_\mu \Psi + \frac i2 \sigma^3  \wt_\mu \cdot \Psi
}
using the Galilean connection $\tilde{\df \w}$ defined via \eqref{eq:comp_structure}. Note that because $\Psi$ is massless, $\nablat_\mu\Psi$ is invariant under $U(1)_M$. As previously discussed, we want to couple the composite fermion to a new gauge field $\df a$ (not to be confused with the mass potential $\df \a$), and so the fully covariant derivative is
\eq{
D_\mu \Psi = \d_\mu \Psi - i a_\mu \Psi + \frac i2 \sigma^3 \wt_\mu  \Psi.
}
Note that any appearance of $D\Psi$ involves coupling to $\df \wt$, not $\df \w$ and we simply drop the tilde for neatness. Since we require a 2+1D Dirac kinetic term, we need an extended tensor whose spatial components are Pauli matrices.
Note that boost invariance tells us that the invariant gamma matrices in the fundamental representation must be
\eq{
\s^A = \begin{pmatrix}
\begin{pmatrix} 0 & 0 \\ 0 & 0 \end{pmatrix} \\
\sigma^1\\
\sigma^2
\end{pmatrix}
}
Unlike our previous discussion in this representation there exists \emph{no} extended tensor invariant under 
\eq{
\Lambda^I{}_J(\Lambda^\dagger)^{-1}\s^J (\Lambda)^{-1}  = \s^I,
}
whose pullback is $\s^A$.  So far we have no time derivatives. However as discussed previously in systems where we have a finite magnetic field turned on we can construct a preferred rest frame and take a time derivative with respect to that,
\eq{
i\Psi^\dagger u^\mu D_\mu \Psi
}
where in order to enforce the lowest Landau level constraint we will use the frame defined by \eqref{eq:_self_cons}, which from here on out we will always assume. 

At which point we may write the kinetic term for the Dirac action to reproduce \eqref{eq:gamma_rep_pauli},\eq{
\frac{i}{2} u^\mu \left( \Psi^\dagger  D_\mu \Psi - D_\mu\Psi^\dagger  \Psi \right) + \frac{i\vf}{2}\left(\Psi^\dagger \s^a D_a \Psi -D_a \Psi^\dagger \s^a \Psi \right) 
}
where $\vf$, the Fermi velocity, is a phenomenological parameter determined by the microscopic physics. Note that, at leading order in derivatives, the time derivative term includes a coupling directly to the electric field,
\eqn{
\frac{iE_a \ce^{ab} }{2B} \Psi^\dagger \overset\leftrightarrow D_b \Psi,
}
which in momentum space is simply $-\ell_B^2( \vec{E}\times \vec{k}) \Psi^\dagger \Psi$. This is simply an energy cost for being in an electric field of the form $\vec{E} \cdot \vec{d}$, an electric dipole moment. The composite Dirac fermion therefore has an electric dipole moment proportional to and orthogonal to its wave-vector, $\vec{d} = \ell_B^2 \hat z \times \vec k$. Note that this is not special to the case of the composite Dirac fermion, as the massless constraint would imply that a time derivative for a single component Schr\"odinger field also has an electric dipole moment of the form $\hat z \times \vec{k}$. Composite fermions having this electric dipole moment has been shown before in other contexts \cite{PhysRevB.47.7312, PhysRevLett.80.4745}.

We can therefore generalize \eqref{eq:son_curved} to satisfy all of our Galilean and LLL constraints with the action
\eqn{\label{eq:action-CDF}
S =& \int |e| \frac{i}{2} \left( u^A \Psi^\dagger \overset \leftrightarrow D_A \Psi +\vf  \Psi^\dagger \s^a \overset \leftrightarrow D_a \Psi \right) \nonumber \\
 &\qquad - \frac{1}{4\pi}\int (\cA+\half \wt)da 
 + \frac{1}{8\pi} \int (\cA+\half \wt)d(\cA+\half\wt)
 }
where we have not included long range Coulomb interactions. Note that in flat space with small electric and magnetic fluctuations about constant $B$,
\eq{
\df \cA + \half \df \wt = \df A  + \frac{ \nabla \cdot E}{4B} dt.
}

\subsection{Conserved currents for the composite fermion}\label{sec:composite-currents}

Following \cite{Geracie:2016dpu} we define the currents
\be
	\ce^\mu = \left. \frac{\delta S}{\delta n_\mu}\right\vert_{\delta \df e^a = \delta \df\alpha = \delta\df\omega^A{}_B = 0},
\quad T^{\mu\nu} = \left. e_A^\nu ~ \frac{\delta S}{\delta \hat e_{A\mu}}\right\vert_{\delta \df n = \delta\df T^I = 0},
\quad s^{\mu\nu\lambda} = \left. e_A^\nu e_B^\lambda ~ \frac{\delta S}{\delta \hat \w_{\mu AB}}\right\vert_{\delta \df e^I = 0}
\ee
where \(\ce^\mu\) is the energy current, \(T^{\mu\nu}\) is the Cauchy stress-mass tensor and \(s^{\mu\nu\lambda}\) is the spin-boost current. In addition, we also have the usual electromagnetic charge current
\be
	J^\mu = \frac{\delta S}{\delta A_\mu}
\ee

On torsionless backgrounds and on-shell, invariance of the action under local Galilean transformations implies that the stress-mass tensor is symmetric. Then the Cauchy stress given by its spatial components and the mass current given by \(\rho^\mu = T^{\mu\nu}n_\nu\) is conserved due to local \(U(1)_M\) invariance. Similarly, diffeomorphism invariance gives us a work-energy equation for \(\ce^\mu\) and a conservation law for \(T^{\mu\nu}\) (see \cite{Geracie:2016dpu} for details).

However, there are further constraints on the currents coming from the form of our action \eqref{eq:action-CDF}. Since, we are working with \(m=0\), \(s=0\) the action is independent of the mass gauge field \(\df\alpha\) and the background spin current \(\df\omega\). As a consequence, we have
\be
	\tilde\rho^\mu = 0, \quad n_\mu s^{\mu\nu\lambda} = 0
\ee
The action does depend on the boost connection \(\df \varpi^a\) through the vorticity \(\Omega\), however  \eqref{eq:LLL_sym} gives us the additional constraint
\be\label{eq:shift-const}
	b^{\mu\nu} = 2 s^{\mu\nu\lambda}n_\lambda = -\tfrac{1}{2} \ce^{\mu\nu} J^0
\ee\\

Now we compute the above currents explicitly about a simple background, in an inertial frame \((t,x,y)\) with constant magnetic field \(B\) and vanishing electric field
\be\label{trivialBackground}
	\df e^I = 
	\begin{pmatrix} 
		dt \\
		d x^i \\
		0
	\end{pmatrix},
	\quad \df T^I = 0 ,\quad \df  A = B x dy
\ee 
We will consider a general statistical gauge field \(\df a\), and decompose it into its magnetic field \(b\) and the electric field \(e_i\).

For this background we have
\be\label{eq:background_flat}
	\tilde{\df T}^I = 0,\quad \tilde{\df\omega}^A{}_B = 0, \quad \df\cA = \df A,\quad u^\mu = (\d_t)^\mu, \quad \Omega = 0
\ee
Note that our chosen inertial frame coincides with the frame of the drift velocity. We will also write down the current components in the above chosen inertial frame, the results in any other frame can be obtained by applying suitable local Galilean transformations (see \cite{Geracie:2015dea, Geracie:2015xfa,Geracie:2016dpu}).

This computation is pretty tedious --- the composite fermion action \eqref{eq:action-CDF} depends explictly on the fields \(u^\mu\), the effective gauge potential \(\df\cA\) and the effective spin connection \(\tilde {\df \w}^A{}_B\) which, in turn, depend on the background fields we need to vary. So we relegate the details to the appendix \ref{sec:currents-calc} and simply summarize the results below.\\

The charge density and currents are given by
\be\label{eq:charge}\begin{split}
	J^t & = \tfrac{1}{4\pi} (B - b) - \d_i P^i \\
	J^i & = - \tfrac{1}{4\pi} \e^{ij}e_j + \d_t P^i \\
\end{split}\ee
where \(P^i\) is the electric dipole moment (or polarization) of the form

\be
	P^i = -\tfrac{i}{2}B^{-1} \e^{ij} \Psi^\dagger \overset\leftrightarrow D_j\Psi -\tfrac{1}{4}\d^i (B^{-1}\Psi^\dagger \sigma^3 \Psi) - \tfrac{1}{16\pi}B^{-1} \d^i (B-b).
\ee
The total charge is $Q = (N_\phi-n_\phi)/2$, and the dipole moment is $\ell_B^2 \hat{z} \times \vec{k}.$\\

The Cauchy mass current and stress are
\be\label{eq:mass-stress}\begin{split}
	\rho^i  & = - \tfrac{1}{4}\e^{ij} \d_j \left[ - \tfrac{1}{2} \Psi^\dagger \sigma^3 \Psi + \tfrac{1}{8\pi} (B - b) - J^t \right]  \\
	T^{ij} & = - \tfrac{i}{2} \vf \Psi^\dagger \s^{(i} {\overset \leftrightarrow D}{}^{j)} \Psi - \tfrac{1}{8\pi} (\d^{(i} e^{j)} - \delta^{ij} \d_k e^k )
\end{split}\ee
Note that the mass density  \emph{locally} vanishes, \(\rho^t = 0 \), while the mass current \(\rho^i\) is pure magentization.\footnote{A similar mass magentization current was found for the Wen-Zee action in \cite{Geracie:2016dpu}.}  \\

The energy density and current is
\be\label{eq:energy}\begin{split}
	\ce^t & = -\tfrac{i}{2} \vf \Psi^\dagger \s^i \overset \leftrightarrow D_i \Psi \\
	\ce^i & = \tfrac{i}{2} \vf \Psi^\dagger \s^i\overset \leftrightarrow D_i\Psi ,
\end{split}\ee
while the boost current is
\be\label{eq:boost-current}
	b^{ij} = -\tfrac{1}{2} \e^{ij} J^t
\ee
satisfying the constraint \eqref{eq:shift-const} from the shift symmetries.\\

Since the action is linear in the statistical gauge field $\df a$, we can integrate it out exactly to find constraints which determine the composite fermion current
\be\begin{split}\label{eq:gauge_constr}
	j^t_{\rm CF} & = \Psi^\dagger\Psi = \tfrac{1}{4\pi}B \\
	j^i_{\rm CF} & = \vf \Psi^\dagger \s^i \Psi = 0
\end{split}\ee

When we turn on electromagnetic perturbations,  the electric dipole moment cancels the leading contribution from the mixed Chern Simons term $ad\wt$, cancelling its contribution of $ \frac{1}{4\pi} \e^{ij} E_j$. In momentum space we find 
\eq{
\delta j^t_{\rm CF} = \frac{\delta B}{4\pi}, ~\delta j^i_{\rm CF}  = \frac{1}{8\pi} \frac{\ell_B^2 \left(  \vec{k}^2\e^{ij}  \delta E^j - \w k^i \delta B\right)}{2+\ell_B^2 \vec{k}^2},
}
and so $j_\mathrm{CF}^ i \sim \d^2 E.$

\section{Conclusion}\label{sec:conclusion}
In this paper we have constructed an action for the composite Dirac fermion that is not merely Galilean invariant but also consistent with additional symmetries of electrons in the lowest Landau level. We demonstrate that \emph{any} low-energy description of electrons in the lowest Landau level feel a nontrivial curvature when electromagnetic fields are turned on. Due to the massless limit of the electrons we find that any composite fermion must have an electric dipole moment $\vec{d} =\ell_B^2 \hat z\times \vec{k}$. Even if we had considered a nonrelativistic composite fermion, a similar argument for constructing a time derivative would again guarantee the dipole moment. We have calculated physical stress, mass and energy currents for this theory as well as the constraint equations imposed by the statistical gauge field.

There are many future directions one can consider following this work: first of all, the proper inclusion of long-range Coulomb interactions in the LLL, following \cite{PhysRevB.95.125120}, has not been considered. We also conjecture that the formalisms outlined here are precisely what is necessary for a covariant completion of the bi-metric theory of FQH states \cite{Gromov:2017qeb}.  Also, now that we have constructed physical operators for the composite Dirac fermion theory, one should be able to directly calculate physical quantities for Jain states such as the Hall viscosity, finite wave-vector Hall conductivity, and static structure factor, and check whether they satisfy various conjectured relations for states in the LLL (some of these checks are addressed in upcoming work \cite{GNRS_to_appear}). It would also be interesting to calculate transport properties for the compressible $\nu =\frac 12$ state itself.

\section*{Acknowledgments}

We thank Dam T. Son, Dung X. Nguyen and Andrey Gromov for insightful discussions and helpful comments on early versions of this paper. This work is supported in part by the NSF grant PHY~15-05124 to the University of Chicago. K.P. is supported in part by the NSF grants PHY-1404105 and PHY-1707800. M.M.R. is supported in part by the DOE grant DE-FG02-13ER41958. 

\appendix

\section{Computation of currents}\label{sec:currents-calc}

To compute the currents we first vary the action \eqref{eq:action-CDF} considering \(\df e^I\), \(\tilde{\df\omega}^A{}_B\), \(u^a\), \(\df\cA\) and \(\df a\) as independent fields (there is no explicit dependence on the mass gauge field \(\df \alpha\))
\be
	\delta S = \int |e| \bigg( - \ce^\mu \delta n_\mu + \hat t^\mu{}_a (\delta e^a)_\mu + \tilde s^{\mu AB} \delta \tilde \omega_{\mu AB} + r_a \delta u^a + \mathcal J^\mu \delta \cA_\mu\bigg)
\ee
to get
\be\begin{split}
	\ce^\mu & = \tfrac{i}{2} \vf \left[ \Psi^\dagger \s^\mu \overset \leftrightarrow D_t\Psi - u^\mu \Psi^\dagger\s^a \overset \leftrightarrow D_a \Psi \right] \\
	\hat t^{\mu\nu} & = - \tfrac{i}{2} \Psi^\dagger \Gamma^\mu \overset \leftrightarrow D{}^\nu \Psi + \tfrac{i}{2} h^{\mu\nu} \Psi^\dagger \Gamma^\lambda \overset \leftrightarrow D_\lambda \Psi\\
	\tilde s^{\mu AB} & = \tfrac{1}{2} \left[ -\tfrac{1}{2}\Psi^\dagger \sigma^3 \Psi u^\mu + \tfrac{1}{8\pi} (B - b) u^\mu - \tfrac{1}{8\pi} \ce^{\mu\nu} e_\nu \right] \e^{AB} \\
	\mathcal J^\mu & = \tfrac{1}{4\pi} (B - b) u^\mu - \tfrac{1}{4\pi} \ce^{\mu\nu}e_\nu \\
	r_a & = \tfrac{i}{2} \Psi^\dagger \overset \leftrightarrow D_a \Psi
\end{split}\ee
where we have used the shorthand \(\Gamma^\mu = u^\mu + \vf \s^\mu \).

Now following \cite{Geracie:2016dpu}, we now convert \(\delta \tilde \omega_{\mu AB}\) into variations of the coframe \(\df e^A\) and the variation of the torsion of the effective connection \eqref{eq:comp_structure}. Note that \(\delta \tilde {\df T}_A = d ( \delta \df \a + \delta u_b \df e^b ) n_A\) on our background. Thus we have
\be\begin{split}
	&\quad \int |e| \left( \hat t^\mu{}_a (\delta e^a)_\mu + \tilde s^{\mu AB} \delta \tilde \omega_{\mu AB} + r_a \delta u^a \right) \\
	& = \int |e| \left( \hat T^\mu{}_a (\delta e^a)_\mu + \hat T^{\mu 0} \delta \alpha_\mu + R_a \delta u^a \right)
\end{split}\ee
with
\be\begin{split}
	\hat T^{\mu 0} & = 0 \\
	\hat T^{\mu \nu} & = \hat t^{\mu \nu} - 2 \nabla_\lambda  \hat S^{\nu \mu \lambda} + 2 \nabla_\lambda (n_\rho \tilde S^{\rho\mu\lambda}) u^\nu \\
	R_a & = r_a + 2 e_{a \mu} \nabla_\lambda (n_\nu \tilde S^{\nu\mu\lambda})
\end{split}\ee
with \(\tilde S^{\lambda \mu \nu} =  \half \left( \tilde s^{\lambda \mu \nu}  - \tilde s^{\mu \nu \lambda} - \tilde s^{\nu \lambda \mu} \right)\).\\

Next we write the variations of \(u^a\) and \(\df\cA\) in terms of variations of \(\df A\) and \(\df \varpi^a\) as follows. For our background \eqref{eq:background_flat}
\be
	\delta \Omega = \ce_{\mu\nu} \nabla^\mu \delta u^\nu + \epsilon^{\mu a} \delta\varpi_{\mu a} 
\ee
so we get 
\be
	\delta \df \cA = \delta \df A - \half (\ce_{\mu\nu} \nabla^\mu \delta u^\nu + \epsilon^{\mu a} \delta\varpi_{\mu a}) \df n
\ee

For the effective drift velocity
\be
	\delta u^\mu = B^{-1} \left( \ce^{\mu\nu\lambda}  - u^\mu \ce^{\nu\lambda} \right) \nabla_\nu \delta \cA_\lambda
\ee
and thus
\be\begin{split}
	R_a \delta u^a & = R_a u^\mu \delta e^a_\mu + R_a e^a_\mu \delta u^\mu \\
		& = R_a u^\mu \delta e^a_\mu + R_a e^a_\mu B^{-1} \ce^{\mu\nu\lambda} \nabla_\nu \delta \cA_\lambda \\
\end{split}\ee

Thus, ignoring the higher derivatives of \(\delta u^\mu\) we have
\be\begin{split}
	&\quad \int |e| \left( \hat T^\mu{}_a (\delta e^a)_\mu + \hat T^{\mu 0} \delta \alpha_\mu + R_a \delta u^a  + \mathcal J^\mu \delta \cA_\mu \right) \\
	& = \int |e| \left( \tilde T^\mu_a (\delta e^a)_\mu + b^{\mu a} \delta \varpi_{\mu a} + J^\mu \delta A_\mu \right)
\end{split}\ee
with
\be\begin{split}
	\tilde T^{\mu\nu} & = \hat T^{\mu\nu} + u^\mu R^\nu \\
	J^\mu & = \mathcal J^\mu + B^{-1}  \ce^{\mu\nu\lambda} \nabla_\nu R_\lambda \\
	b^{\mu\nu} & = - \tfrac{1}{2} J^0 \ce^{\mu\nu} \\
\end{split}\ee

Finally we carry out the ``improvement'' procedure in \cite{Geracie:2016dpu} to get the Cauchy stress-mass tensor
\be\begin{split}
	T^{\mu \nu} & = \tilde T^{\mu \nu} - 2 \nabla_\lambda (-\tfrac{1}{4} J^0 u^\nu \ce^{\mu\lambda} )
\end{split}\ee

Explicitly, computing all the above quantities we get the currents summarized in Sec.\ref{sec:composite-currents}.

\addcontentsline{toc}{section}{Bibliography}
\bibliographystyle{JHEP}
\bibliography{NR_refs}

\end{document}